\documentclass[reqno]{amsart}

\usepackage{amssymb}
\usepackage{graphicx}
\usepackage{amsfonts}
\usepackage{bm}

\newcommand{\bee}{\begin{eqnarray}}
\newcommand{\eee}{\end{eqnarray}}
\newcommand{\be}{\begin{eqnarray*}}
\newcommand{\ee}{\end{eqnarray*}}
\newcommand{\R}{{\mathbb R}}

\newcommand{\A}{A}
\newcommand{\E}{z}

\begin{document}

\title [Resonances crossings] {Resonances crossing effect and quantum sensor of electric fields}
 
 \author {
 Andrea Sacchetti
 }

\address {
Department of Physics, Informatics and Mathematics, University of Modena and Reggio Emilia, Modena, Italy.
}

\email {andrea.sacchetti@unimore.it}

\date {\today}

\thanks {Andrea Sacchetti is member of Gruppo Nazionale per la Fisica Matematica of Istituto Nazionale di Alta Matematica (GNFM-INdAM). \ This work is partially supported by the Next Generation EU - Prin 2022CHELC7 project "Singular Interactions and Effective Models in Mathematical Physics"
 and the UniMoRe-FIM project  ``Modelli e metodi della Fisica Matematica''.}

\begin {abstract} 

While the phenomenon of the exact crossing of energy levels is a rarely occurring event, in the case of quantum resonances associated with metastable states this phenomenon is much more frequent and various scenarios can occur. \ When there is an exact crossing of the imaginary parts of the resonances in a two-level quantum system subject to an external DC electric field, then a damped beating phenomenon occurs, which is absent if there is no such crossing. \ This fact, tested numerically on an explicit one-dimensional model, suggests the possibility of designing quantum sensors to determine in a very simple way whether the external field strength has an assigned value or not.

\bigskip

{\it Data availability statement.} \ All data generated or analysed during this study are included in this published article.

\bigskip

{\it Conflict of interest statement.} \ The author has no competing interests to declare that are relevant to the content of this article.
\end{abstract}



\maketitle

{\bf Introduction.} \ Quantum sensors are devices that measure physical quantities, such as electric or magnetic fields, using the laws of Quantum Mechanics. \ Among the first devices with these characteristics we can mention the Ramsey interferometer \cite {Ramsey}, based on a two-level system. \ Currently, many other quantum sensors have been developed, from those based on silicon carbide to those based on NV centers in nanodiamonds, with many applications, from rechargeable batteries to biomedical devices (see \cite {Dolde,Aslam,Hoo,Castelletto} and references therein). 

The typical model of a quantum sensor \cite {Degen} consists of an Hamiltonian $ H = H_i + V_e$, where $H_i =$ is the internal (or unperturbed) Hamiltonian, and $V_e$ is an external potential that one wants to measure or that represents an element of control used to set the quantum sensor appropriately; for example, in some quantum sensors the external field is a DC electric field, and $V_e(x)$ represents a linear Stark potential. \ Assume that $H_i$ is a two-level system with energy levels $E_1\le E_2$, and term $\omega =E_2-E_1$ is the transition frequency between the two states; therefore, in the case where $\omega \not= 0$, an interference effect generates a periodic beating effect with period $T=2\pi /\omega$. \ The basic idea is that the effect of the perturbation dues to $V_e$ is to produce a shift in energy levels, and a change in the transition frequency $\omega$; so, a periodic beating motion of the quantum system described by $H$ is still observed, but with modified period. \ A relationship is then established between the period of the beating motion and the external field that can be used to obtain an estimate of the intensity of the perturbation  $V_e$. 

In this paper, by making use of a similar idea we propose a theoretical model for a quantum sensor aimed at determining in a simple way whether or not an external DC electric field strength is equal to a predefined value. \ In some ways we have a phenomenological analogy with RLC electrical circuits that resonate at a specific external frequency; in our model, similarly, only at a specific value of the DC electric field strength a long time interference between two metastable states producing a periodic beating effect occurs.

{\bf Interference effect at the resonances crossing point.} \  Let the internal one-dimensional Hamiltonian $H_i=-\frac {d^2}{dx^2} + V_i$ be associated with an asymmetrical double-well potential $V_i$, and let the external potential be a linear Stark potential $V_e (x) = - F x$,  where $F$ is the DC electric field strength. \ If we consider the ground state of a single well, {\it e.g.} the left-hand-side (l.h.s.) one, the effect of the second well is to slightly perturb it, and we denote by $E_1$ the corresponding energy; similarly, let us denote by $E_2$ the energy of the ground state of the right-hand-side (r.h.s.) well treating the l.h.s. well as a perturbation. \ The two-level system consists by restricting $H_i$ to these two states, and the effect of the linear Stark potential $V_e$ is, in the first instance, to change the splitting $\omega$ as described above. \ We can observe, however, that a second effect, associated with crossing energy levels, is generated for an appropriate choice of parameters. \ Indeed, if the l.h.s. well of $V_i$ is deeper than the r.h.s. well, then, for $F=0$, the energy level $E_1$ is less than the energy level $E_2$. 

\begin{figure}
\begin{center}
\includegraphics[height=6cm,width=8cm]{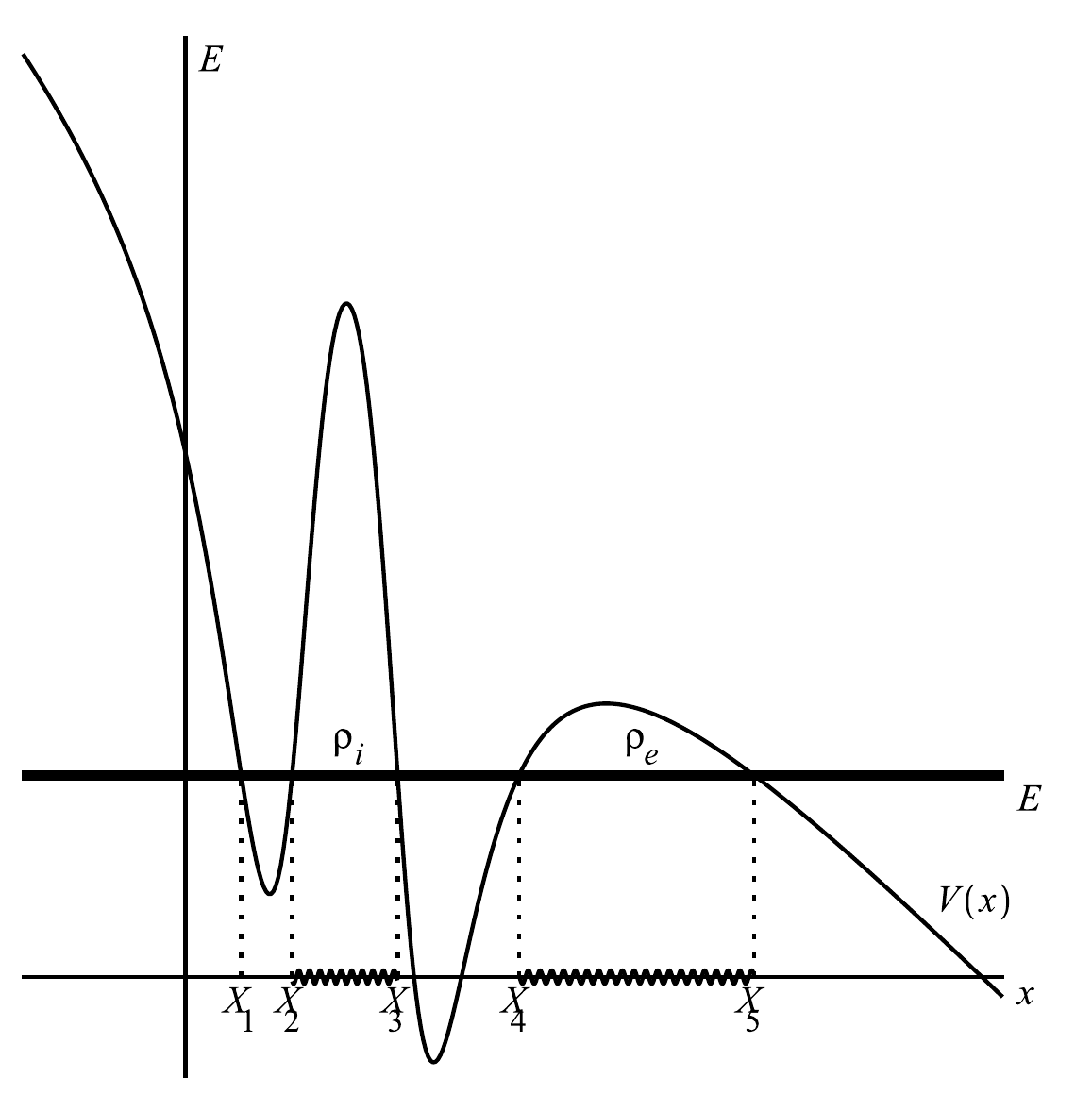}
\caption{Asymmetrical double-well model with potential $V=V_i+V_e$; the inner barrier with Agmon length $\rho_i$ is between $X_2$ and $X_3$, and the outer barrier with Agmon length $\rho_e$ is between $X_4$ and $X_5$. \ There are no stable states for $H$, but metastable ones.}
\label {Barriera}
\end{center}
\end{figure}

As $F$ increases, these two levels will come closer together until they almost cross each other for a critical value of $F$, and we denote by $E$ their (almost) common value.  \  Since in dimension one energy levels are always non-degenerate then exact crossing is not possible, and thus we have an avoided-crossing picture. \ At this  (almost) common value $E$ of the two levels at the avoided-crossing point the two wells are separated by an inner barrier, where the "classical motion" is forbidden since the potential $V(x) = V_i (x) + V_e (x)$ exceeds the energy level $E$; also there is an outer barrier (see fig. \ref {Barriera}). 

\begin{figure}
\begin{center}
\includegraphics[height=4cm,width=4cm]{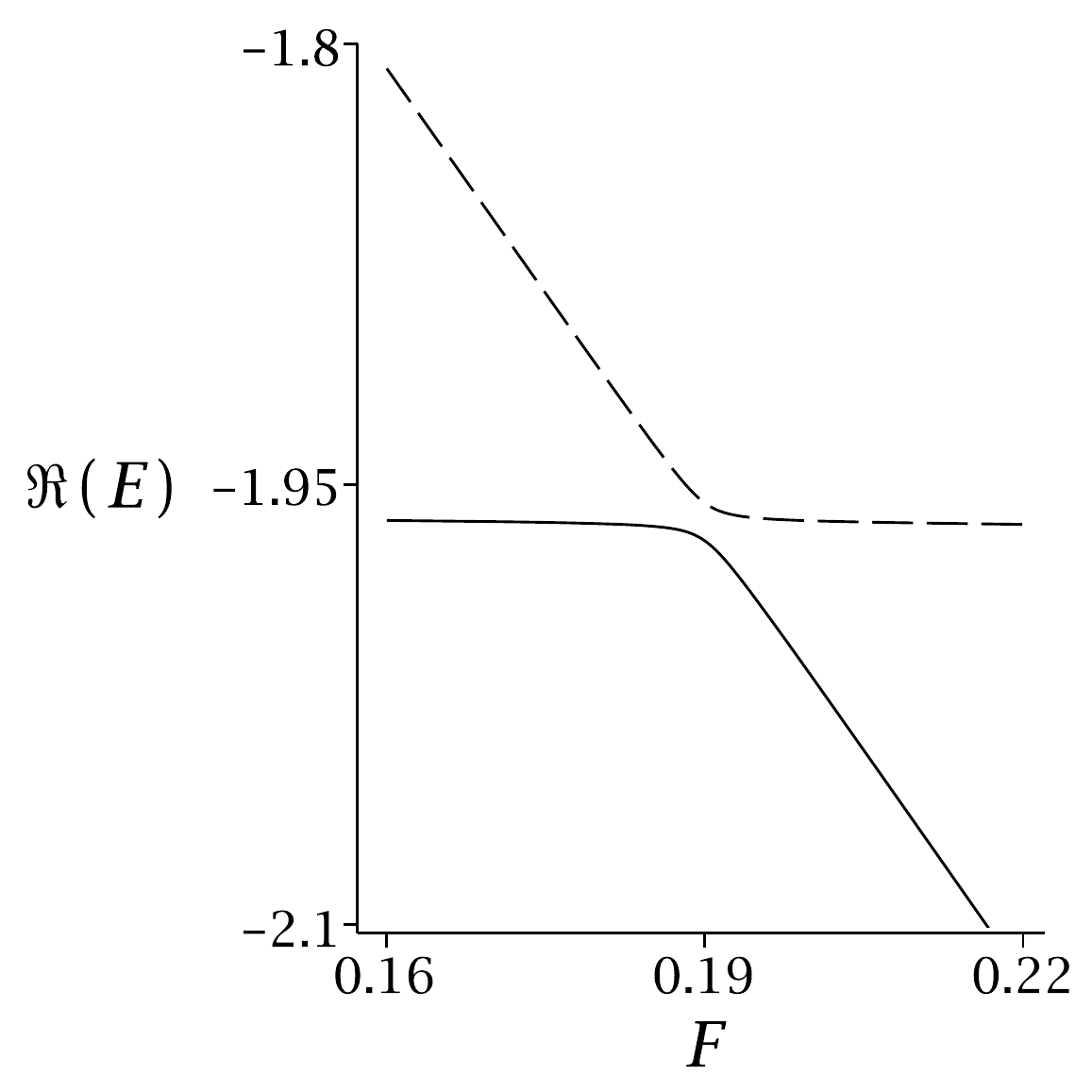}
\includegraphics[height=4cm,width=4cm]{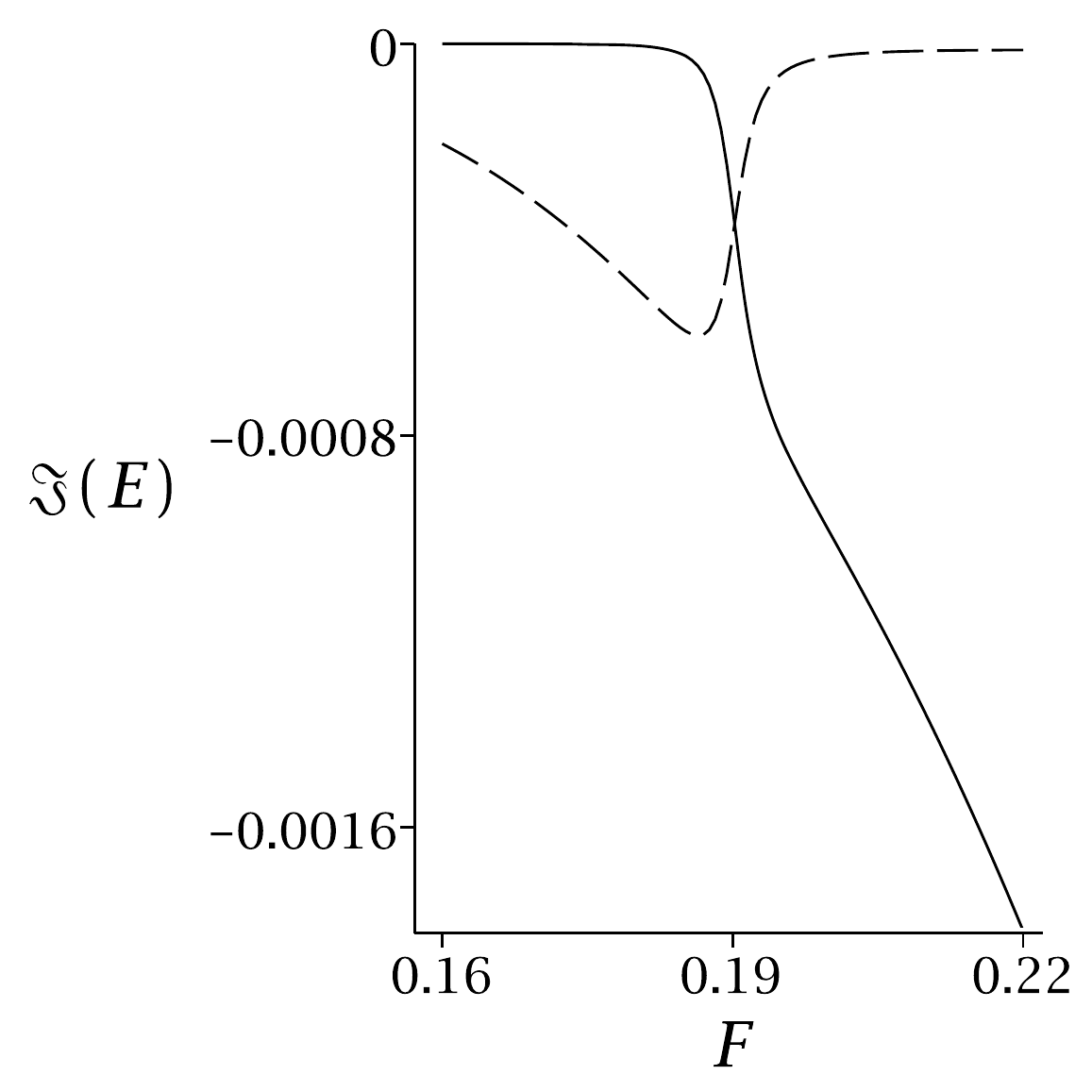}
\caption{Type I crossing for $a=5$, $\alpha_1 = -2.8$, and $\alpha_2 = -2$. \ In the l.h.s. panel we plot the real part of $E_1 (F)$ (full line) and $E_2 (F)$ (broken line); in the r.h.s. panel we plot the imaginary part of $E_1 (F)$ (full line) and $E_2 (F)$ (broken line).}
\label {CrossI}
\end{center}
\end{figure}

\begin{figure}
\begin{center}
\includegraphics[height=4cm,width=4cm]{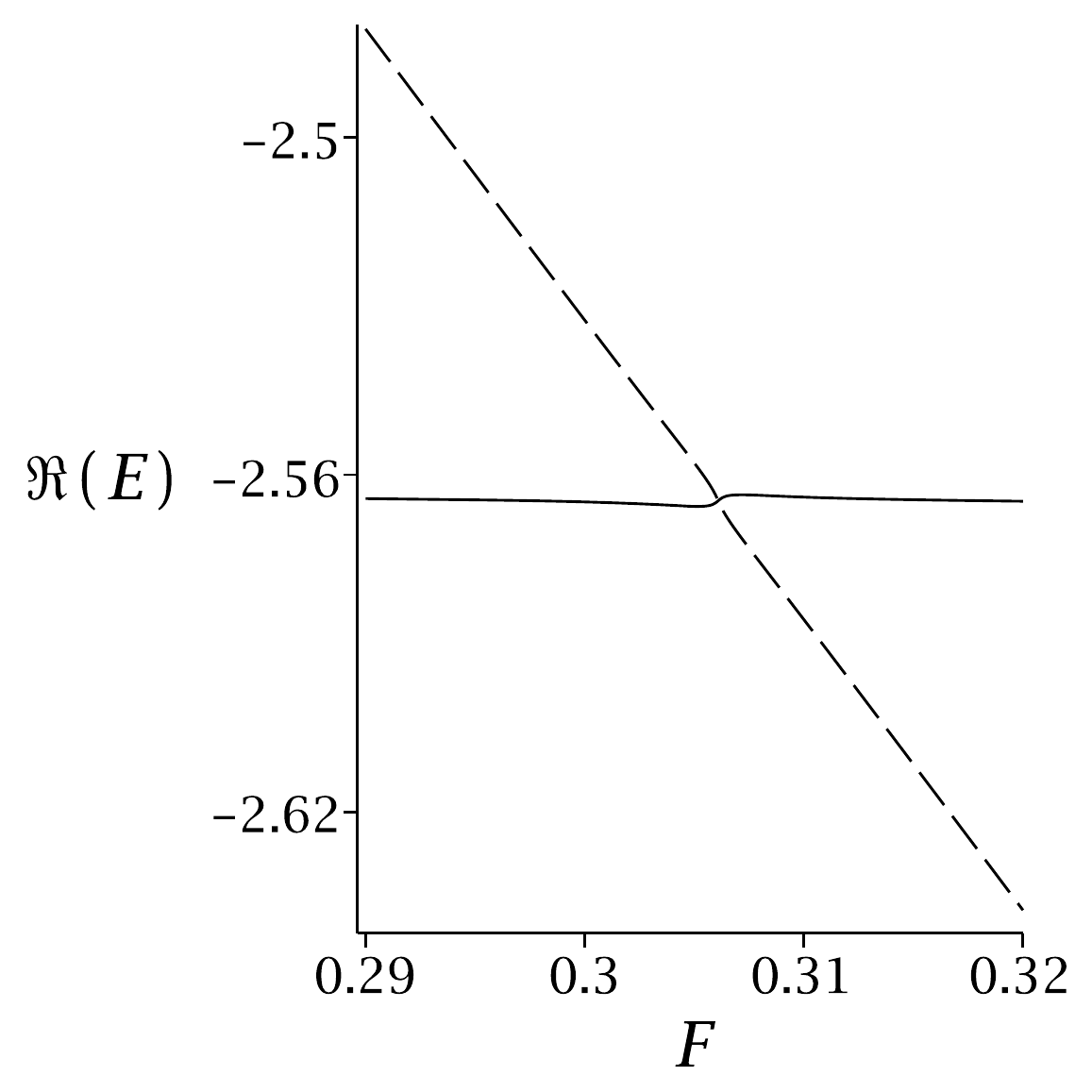}
\includegraphics[height=4cm,width=4cm]{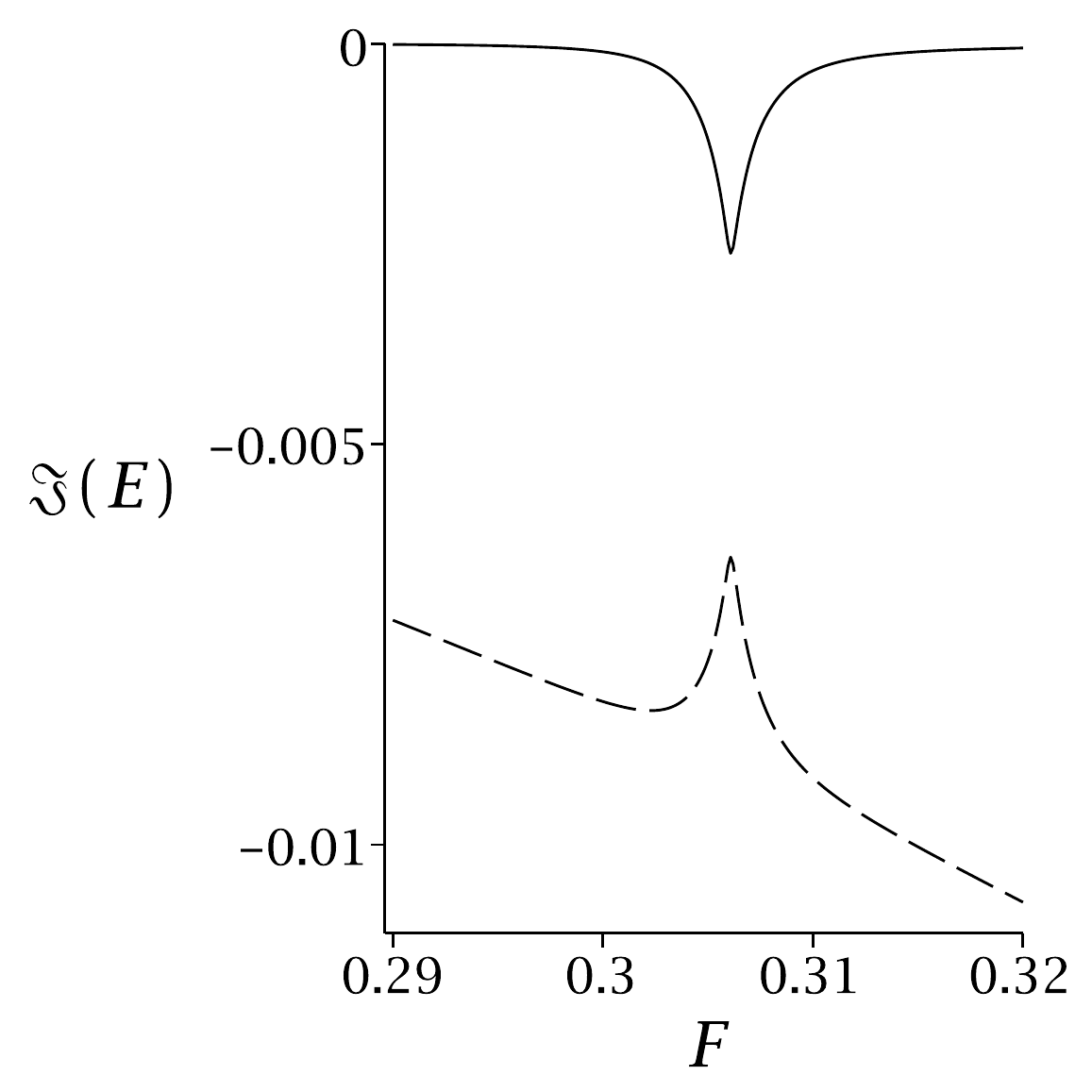}
\caption{Type II crossing for $a=5$, $\alpha_1 = -3.2$, and $\alpha_2 = -2$. \ In the l.h.s. panel we plot the real part of $E_1 (F)$ (full line) and $E_2 (F)$ (broken line); in the r.h.s. panel we plot the imaginary part of $E_1 (F)$ (full line) and $E_2 (F)$ (broken line).}
\label {CrossII}
\end{center}
\end{figure}

We can measure the length of these barriers by means of the Agmon metric  \cite {Agmon} that in dimension one is simply $\rho_i = \int_{X_2}^{X_3} \sqrt {V(x)-E} \, dx $, for the inner barrier, and  
$ \rho_e = \int_{X_4}^{X_5} \sqrt {V(x)-E} \, dx $ that of the outer barrier.
In fact, for any $F>0$ the stationary states become metastable states associated to quantum resonances $E_1 (F)$ and $E_2 (F)$ depending on $F$ and with negative imaginary part, and in the complex plane the crossing phenomenon  between these two levels may be of:
\begin {itemize} 

\item [-] type I crossing when there is an exact crossing of their imaginary parts $\Im E_1 (F)$ and $\Im E_2 (F)$, and an avoided-crossing of their real parts $\Re E_1 (F)$ and $\Re E_2 (F)$ (see, {\it e.g.},  fig. \ref {CrossI}), which occurs when $\rho_i < 2 \rho_e$;

\item [-] type II crossing when there is an exact crossing of their real parts, and an avoided-crossing of their imaginary parts (see, {\it e.g.},  fig. \ref {CrossII}), which occurs when $\rho_i > 2 \rho_e$.

\end {itemize}

Crossing phenomenon of resonances was analyzed as early as Avron's work in 1982 \cite {Avron},  and was later taken up by \cite {GS,GMS1,GMS2} providing the above criterion for determining the type of crossing. \ This theoretical result is valid in the semiclassical limit, which in the present context means that the parameters $\alpha_1$, $\alpha_2$ and $F$ are very large in absolute value so that the two resonances are very narrow, that is their imaginary parts are very small in absolute value. \ However, from a practical point of view, we will see in a numerical experiment that this criterion is already useful even if the parameters are not excessively large. \ We should mention that the question of resonances crossing is of great interest both in a general context \cite {KKM,HJM}, and even in  experimental applications \cite {XWBH, WBXH}; furthermore, it has also been analyzed in the case of symmetrical double-well potential \cite {Korsch}. 

When the imaginary parts of the two resonances $E_1 (F)$ and $E_2 (F)$ are sufficiently different from each other then the lifetime of the two metastable states are quite different, and the phenomenon of interference will not be triggered because one of the two metastable states decays much faster than the other one. \ In fact, a very slight beating effect can be observed for small times, as pointed out by \cite {Jung}, but it quickly disappears and only the exponential decay associated with the narrowest resonance remains as the dominant term. \ On the other side, this picture changes markedly at values of model parameters for which the imaginary parts of the two resonances coincide $\Im E_1 (F) = \Im E_2 (F)$; in this case, which can occur only in the case of type I crossing, the interference effect is triggered, and thus we observe a damped beating effect with (pseudo-)period $T=2\pi /\left ( \Re (E_2) - \Re (E_1 ) \right )$ for large time intervals. 

In this way, we have a theoretical model for a quantum sensor that makes it possible to verify whether  the intensity $F$ of the external field assumes the specific value for which type I crossing and beating effect occur; we observe that this specific value can be appropriately selected by adjusting the value of the parameters of the internal double-well potential $V_i$. 

{\bf Qualitative analysis in a simple model.} To qualitatively verify this idea on an explicit one-dimensional model, we consider the double-well potential with two attractive singular interactions due to two Dirac's $\delta$,
\be
V_i (x) = \alpha_1 \delta_{x_1} + \alpha_2 \delta_{x_2} \, ,  \ x_1 < x_2  \, , \ \alpha_1 <\alpha_2 <0 \, , 
\ee
where $\delta_{y}$ is such that $\int_{\R} \delta_{y} f(x) dx = f(y)$. \ The idea of using Dirac's $\delta$ to model  wells or barriers goes back to Enrico Fermi \cite {Fermi}, and in this context it has been extensively used for the study of the Stark effect on both a single well \cite {Lukes, Ludviksson,Nickel,Emmanouilidou,Alvarez1,Rokhlenko}, and on double-well \cite {Korsch,Alvarez2,Jung,HM} cases.

The spectrum of $H$ is purely continuous and it coincides with the entire real axis. \ Therefore, no stable states are possible. \ It is possible, however, to have resonances $E$ associated with metastable states $\psi$ such that $H\psi = E\psi$ where $\psi$ satisfies the outgoing conditions \cite {Korsch} $ \psi (x)= C_i^+ (x) $ for $x >a$, 
where $C_i^+ (x)=B_i (x)+i A_i(x)$, $A_i$ and $B_i$ being the two Airy functions. \ Resonances of $H$ can be equivalently defined as the complex poles of the analytic continuation of the kernel of the resolvent operator $[H-\E ]^{-1}$ from the upper half-plane $\Im \E >0$ to the lower half-plane $\Im \E <0$ (see \cite {HM} and references therein). \ If we denote by $K_0= K_0^\pm$, if $\pm \Im \E >0$, the kernel of the resolvent operator $[H_0-z]^{-1}$, where $H_0 = - \frac {d^2}{dx^2} - Fx$,  then it is known that
\be
K_0^{\pm } (x,y;\E ) = \frac {\pi}{F^{1/3}} 
\left \{ 
\begin {array}{l}
 Ci^\pm \left ( - \frac {F y + {\E }}{F^{2/3}} \right )  
 Ai \left (- \frac {F x + {\E }}{F^{2/3}}  \right )\, , x \le y \\ 
  Ci^\pm \left (- \frac {F x + {\E }}{F^{2/3}}  \right )  A_i
  \left (- \frac {F y + {\E }}{F^{2/3}} \right ) \, ,y < x 
\end {array}
\right. 
.
\ee
The kernel $K = K^\pm$, if $\pm \Im \E >0$, of the resolvent operator $[H-z]^{-1}$ can be obtained by $K_0^\pm$ as follows \cite {Na}:
\be
 K^\pm (x,y;\E ) = K_0^\pm (x,y;\E ) +  \frac {R^\pm (x,y;\E )}{D^\pm (\E )}\, , 
 \ee
 where
 \be 
 R^\pm (x,y;\E )= \sum_{n,m=1}^2  K_0^\pm (x,x_n;\E ) M^\pm_{n,m} (\E )  K_0^\pm (x_m,y;\E )
\, ,
\ee
 and where, adopting the shortcut notation $k^\pm_{j,\ell} (\E )=  K_0^\pm (x_j,x_\ell ;\E ) $,
\be
D^\pm (\E ) =    \frac {\left [ {1}+{\alpha_1}  k^\pm_{1,1} (\E ) \right ]\,  \left [  {1}+{\alpha_2}  k^\pm_{2,2} (\E ) \right ]}{\alpha_1 \alpha_2}-  k^\pm_{1,2}(\E )  k^\pm_{2,1}(\E ) \, , 
\ee
and
\be
M^\pm(\E ) := \left ( 
\begin {array}{cc}
\frac {1}{\alpha_2} + k^\pm_{2,2}(\E ) & - k^\pm_{1,2}(\E ) \\
- k^\pm_{2,1}(\E ) &  \frac {1}{\alpha_1} + k^\pm_{1,1}(\E )  
\end {array}
\right )\, . 
\ee
Then resonances $\E$ for $H$ are the zeros such that $\Im z <0$ of the analytic continuation of the function $D^+ (\E )$ from the upper half-plane $\Im \E >0$ to the lower half-plane $\Im \E <0$, and it is possible to compute them numerically once parameter values $F$, $\alpha_1$, $\alpha_2$, $x_1$ and $x_2$ are assigned.

The two resonances have respectively real parts close to the values of the two single-well ground states: $\Re E_1 \sim - \alpha_1^2/4$ and $\Re E_2 \sim - \alpha_2^2/4 - Fa $, where $\sim$ means the asymptotic value for large $|\alpha_{1,2}|$, and where we set, to fix the ideas and without losing in generality, $x_1=0$ and $x_2 =a>0$. \ Thus crossing phenomenon occurs when $F$ is close to the critical value
\bee
F_C \sim \frac {\alpha_1^2 - \alpha_2^2}{4a} \, , \label {Semi}
\eee
and for these values we have that the Agmon lengths of the inner and outer barriers are given by $\rho_i \sim 
(\alpha_1^3-\alpha_2^2)/{12 F_C}$ and $\rho_e \sim  {\alpha_2^3}/{12 F_C}$. \ Thus a type I crossing occurs when $\sqrt[3]{3} \gtrsim \alpha_1/\alpha_2$, and type II when $\alpha_1/\alpha_2 \gtrsim \sqrt[3]{3}$. \ Actually, this theoretical result, which holds true exactly only in the semiclassical limit in which $| \alpha_{1,2}|$ and $|F|$ go to infinity, retains good validity even when the parameters are finite. \ For example, for $a=5$, $\alpha_1 =-2.8$, and $\alpha_2 =-2$, a type I crossing is observed in fig. \ref {CrossI}; from the semiclassical result (\ref {Semi}), it turns out that the theoretical value of $F_C$ is $0.192$, and this result is in good agreement with the numerical experiment in which $F_C$ takes the value of $0.190(2)$. \ For $a=5$, $\alpha_1 =-3.2$, and $\alpha_2 =-2$ a type II cross is observed in fig. \ref {CrossII}, and again there is good agreement between (\ref {Semi}) and the result of the numerical experiment. 

An important remark concerns the fact that from the type of crossing follows a different behaviour of observables. \ Let us consider, for example, the survival amplitude $\A (t)= \langle \psi_0 | \psi_t \rangle $ where $\psi_0$ is the wave function of the initial state, and where $\psi_t = e^{-iHt}\psi_0$ is the wave function of the state at instant $t$. \ In absence of stable states the survival amplitude decreases in time \cite {HM}. \ This decay is, for a generic Hamiltonian without stable states, the contribution of two terms: one of exponential type that is dominant for intermediate time intervals, and one of power type that instead becomes dominant for longer times; see  \cite {Wi} where it was numerically conjectured that a transition effect between the two different types of decay starts around a certain instant $t$, see also \cite {AgSa1,de,E,GMV,GL,GVHR,PVZ,RaK} for rigorous results, and  \cite {W} for the experimental evidence of the deviation from exponential decay. \ However, in our model the power law decay does not play any role because the spectrum of $H$ is not bounded from below \cite {Simon}, and thus we expect to observe only the exponential decay. \ In particular,  since there are only two narrow resonances $E_1 (F)$ and $E_2 (F)$ having small imaginary part then from the residue Theorem it follows that the dominant term of the survival amplitude is given by \cite {Nickel}
\be
\langle \psi_0 | e^{-itH_0}\psi_0 \rangle + \sum_{j=1}^2 c_j e^{-it E_j (F)} \, , 
\ee
where
\be
c_j = R_j \sum_{n,m=1}^2 M^+_{n,m} \bm {(} E_j (F) \bm {)}\,  q_{n,j} p_{m,j}\, , 
\ee
and
\be
q_{n,j} &=& \int_{\R} K_0^+ \bm {(} x,x_n;E_j(F)  \bm {)}\,  \bar \psi_0 (x) dx 
\\ 
p_{m,j} &=& \int_{\R} K_0^+ \bm {(} x_m,y;E_j(F)  \bm {)}\,   \psi_0 (y) dy
\ee
and where $R_j$ is the value of the residue of the function $1/D^+(\E )$ at $\E = E_j (F)$. \ The evolution operator $e^{-itH_0}$ is an integral operator with kernel \cite {AH} 
\be
U_0 (x,y;t )
= \frac {\mbox {exp} \left \{  \frac {i}{2} \left [ \frac {(x-y)^2}{2t} -\frac {1}{6} F^2 t^3 +{ t F (x+y)}   \right ] \right \}}{\sqrt {4\pi i t}} 
\, ,
\ee
and the term $\langle \psi_0 | e^{-itH_0}\psi_0 \rangle$ decays more rapidly than the exponential $e^{-t |\Im E_j (F)|}$ in the case of narrow resonances. \ Then, the asymptotic behaviour of $\A (t)$ is governed only by the contributions given by the two resonances. \ If we are in the type II crossing case, where {\it e.g.} $|\Im E_1 (F)| < |\Im E_2 (F)|$ for each value of $F$ as in fig. \ref {CrossI}, then the dominant contribution to the exponentially decreasing behaviour of $|\A (t) |$ is given by $ |c_1| e^{-t |\Im E_1 (F) |}$. \ If, on the other hand, we are in the case of type I crossing, and if we call $F_C$ the value of $F$ at which $\Im E_1 (F) = \Im E_2 (F)$ then we still observe an 
exponentially decreasing behaviour given by $ |c_j| e^{-t |\Im E_j (F) |} $, where $j=1$ if $F <F_C$, and $j=2$ if $F>F_C$ as in fig. \ref {CrossII}; eventually, only for $F$ close to $F_C$ there is a damped oscillating behaviour due to the interference between the two resonances, and in this case the dominant behaviour of $|\A (t) |$ is given by:
\be
e^{- t |\Im E_1 (F_C )|} \left | c_1 + c_2  e^{-i \omega t } \right ]\, , 
\ee
with pseudo-period $T= {2\pi}/{\omega}$ where $\omega = \Re E_2 (F_C)-\Re E_1 (F_C)$. 

For $a=5$, $\alpha_1 =-2.8$, and $\alpha_2 =-2$, we have a type I crossing when $F$ takes the value $F_C =0.190(2)$. \ We consider a numerical experiment where $\psi_0$ is a normalized Gaussian localized on the l.h.s. well corresponding to $x=0$:
\be
\psi_0 (x) = (2 \pi \sigma^2)^{-1/4} e^{-x^2/4\sigma^2}\, ,\ \mbox { where  } \sigma =1/2\, , 
\ee
and we go on to study the behaviour of $\A (t)$ for $t \in [10^2,10^5]$ and for some $F<F_C$, $F>F_C$, and for $F=F_C$. \ For the values of the resonances $E_{j}$, and of the coefficients $c_j$, $j=1,2$, see table \ref {tabella1}.
\begin{table}
\begin{center}
\begin{tabular}{lcccccc} 
\hline \hline 
          &\  & $F=0.17 < F_C$   &    \     \       & $F= F_C = 0.190(2)$ & \ & $F= 0.21 > F_C$ \\  \hline 
$\Re E_1$ &\ & $ -1.96(3) $            &  \        & $ -1.97(0) $  &\ & $ -2.06(6)$  \\ 
$\Im E_1$ &\ & $ -0.35(1)\cdot 10^{-8} $  &\ & $-0.36(8) \cdot 10^{-3}$ &\ & $ -0.13(7) \cdot 10^{-2} $  \\ 
$\Re E_2$ &\ & $ -1.86(0)$                    &\   & $ -1.95(7)$  &\ & $ -1.96(3)$   \\ 
$\Im E_2 $ &\ & $ -0.32(9)\cdot 10^{-3}$ &\  & $ - 0.36(8) \cdot 10^{-3}$  &\ & $  -0.14(6) \cdot 10^{-4} $   \\ 
$\Re c_1$  &\ & $  0.41(0)\cdot 10 ^{-2}$ &\ & $ 0.44(8) $  &\ & $ 0.96(5)$  \\ 
$\Im c_1$  &\ & $   0.85(1)\cdot 10^{-7}$ &\ & $- 0.027(6)$  &\ & $ 0.13(0) \cdot 10^{-3} $  \\ 
$\Re c_2$ &\ & $ 0.96(6) $ &\ & $0.52(1)$ &\  & $0.34(9) \cdot 10^{-2}$  \\ 
$\Im c_2$  &\ & $ - 0.11(2)\cdot 10^{-4}$ &\ & $0.027(5)$  &\ & $- 0.14(4) \cdot 10^{-3} $  \\ \hline  \hline 
\end{tabular} 
\caption{Table of values of the resonances $E_j$ and of the coefficients $c_j$, $j=1,2$, for different values of $F$; where $a=5$, $\alpha_1 =-2.8$, and $\alpha_2 =-2$ correspond to a type I crossing.}
\label{tabella1}
\end{center}
\end {table}
\begin{figure}
\begin{center}
\includegraphics[height=6cm,width=8cm]{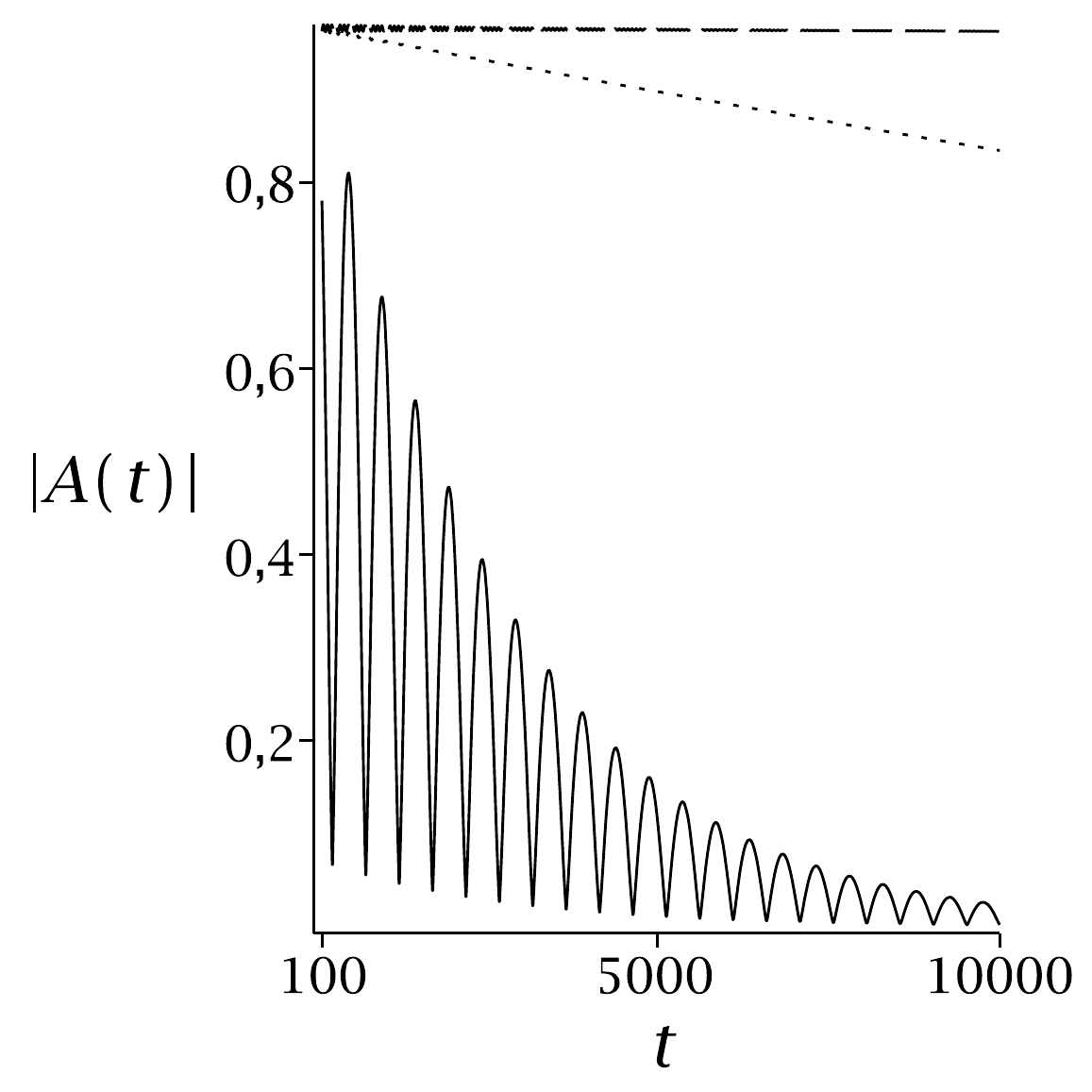}
\caption{In the case of type I crossing corresponding to $a=5$, $\alpha_1 = -2.8$, and $\alpha_2 = -2$, we can observe in the interval $t \in [10^2,10^4]$, a damped beating behaviour (full line) of $|\A (t) |$ when $F$ is close to the critical value $F_C =0.190(2)$ for which we have exact crossing of the imaginary parts of the two resonances;  if $F =0.17 <F_C$ (broken line) or $F =0.21 >F_C$ (dot line) no significant oscillation is observed.}
\label {Battimenti}
\end{center}
\end{figure}
As expected, in the l.h.s. picture of fig. \ref {Battimenti} a damped oscillating behaviour is observed for $|\A (t) |$ when the intensity $F$ of the external field is close to $F_C$. \ On the other hand, when $F$ is different from this value a damping without significant oscillations is observed in the r.h.s. picture of fig. \ref {Battimenti} because one of the two metastable states decays much faster than the other one, and therefore it is not possible to have a long times interference phenomenon. \ In fact, oscillations with small amplitude can be seen only for short times and the damping effect is slower than in the first case, because one of the two resonances is much narrower than those obtained when $F=F_C$. \ We remark that the critical value $F_C$ does not depend on the initial state; instead, as the parameters $\alpha_1$, $\alpha_2$ and $a$ vary, one can tune the critical value $F_C$ as desired.

{\bf Conclusions.} \ In conclusion, time evolution of the survival amplitude for a Schr\"odinger operator with asymmetrical double-well potential under the effect of a Stark perturbation is analyzed in detail in this paper. \ It is verified that it is a theoretical model useful to design a quantum sensor for which the response to an external DC electric field has two distinctly different behaviours depending on whether the intensity of the field is close to or different from a predetermined value $F_C$ of the external field strength. \ By means of (\ref {Semi}) the  value $F_C$  can be chosen by tuning the parameters of the internal potential. \ The novelty is that it is possible to verify a specific value of the DC electric field simply by observing whether or not significant oscillations are present, no matter their period.


\begin{thebibliography}{99}

\bibitem {Ramsey} N.F. Ramsey, 
%
%
Phys. Rev. {\bf 78} 695 
(1950).

\bibitem {Dolde} F. Dolde, H. Fedder, M.W. Doherty, T. 
N\"obauer, F. Rempp, G. Balasubramanian, T. Wolf,
F. Reinhard, L.C.L. Hollenberg, F. Jelezko, and J. Wrachtrup,
Nature Phys. {\bf 7} 459 
(2011).

\bibitem {Aslam} N. Aslam, H. Zhou, E.K. Urbach, M.J. 
Turner, R.L. Walsworth, M.D. Lukin, and H. Park, 
Nat. Rev. Phys. {\bf 5} 157 
(2023).

\bibitem {Hoo} M. Hollendonner, S. Sharma, S.K. Parthasarathy, D.B.R. Dasari, A. Finkler, S.V. Kusminskiy, and R. Nagy, 
New J. Phys. {\bf 25} 093008 
(2023).

\bibitem {Castelletto} S. Castelletto, C.T.-K. Lew, W.-X. 
Lin, and J.-S. Xu, 
Rep. Prog. Phys. {\bf 87} 014501 
(2024).

\bibitem {Degen} C.L. Degen, F. Reinhard, and P. Cappellaro, 
Rev. Mod. Phys. {\bf 89} 035002 
(2017).

\bibitem {Agmon} S. Agmon, {\it Lectures on exponential decay of solutions of second-order elliptic equations: bounds on eigenfunctions of N-body Schr\"odinger operators}. Mathematical Notes, {\bf 29}. Princeton University Press, 1982.

\bibitem {Avron}  J.E. Avron, Ann. Phys. (NY) {\bf 143} 33 
(1982).

\bibitem {GS} V. Grecchi, and A. Sacchetti, Ann. Phys. (NY) {\bf 241} 258 
(1995).

\bibitem {GMS1} V. Grecchi, A. Martinez, and A. Sacchetti, Asympt. Anal. {\bf 13} 373 
(1996).

\bibitem {GMS2} V. Grecchi, A. Martinez, and A. Sacchetti, J. Phys. A: Math. Gen. {\bf 29} 4561 
(1996).

\bibitem {KKM} F. Keck, H.J. Korsch, and S. Mossmann, J. Phys. A: Math. Gen. {\bf 36} 2125 
(2003).

\bibitem {HJM} E. Hern\'andez, A. J\'auregui, and A. Mondrag\'on, Int. J. Thoer. Phys. {\bf 46} 1890 
(2007).

\bibitem {XWBH} Y. Xia, L. Wang, D. Bai, and W. Ho, ACS Nano {\bf 17} 23144 
(2023).

\bibitem {WBXH} L. Wang, D. Bai, Y. Xia,  and W. Ho, Phys. Rev. Lett. {\bf 130} 096201 
(2023).

\bibitem {Korsch} H.J. Korsch, and S. Mossmann, J. Phys. A: Math. Gen. {\bf 36} 2139 
(2003).

\bibitem {Jung} J.-W. Jung, K. Na, and L.E. Reichl, Phys. Rev. A {\bf 80} 012518 
(2009).

\bibitem {Fermi} E. Fermi, Ric. Sc. {\bf 7} (2), 13 
(1936).

\bibitem {Lukes} T. Lukes, and K.T.S. Somaratna, J. Phys. C: Solid State Phys. {\bf 2} 586 
(1969).

\bibitem {Ludviksson} A. Ludviksson, J. Phys. A: Math. Gen. {\bf 20} 4733 
(1987).

\bibitem {Nickel} J.C. Nickel, and L.E. Reichl, Phys. Rev. A {\bf 58} 4210 
(1998).

\bibitem {Emmanouilidou} A. Emmanouilidou, and L.E. Reichl, Phys. Rev. A {\bf 62} 022709 
(2000).

\bibitem {Alvarez1} G. \'Alvarez, and B. Sundaram, Phys. Rev. A {\bf 68} 013407 
(2003).

\bibitem {Rokhlenko} A. Rokhlenko, Phys. Rev. A {\bf 78} 022113 
(2008).

\bibitem {Alvarez2} G. \'Alvarez, and B. Sundaram, J. Phys. A: Math. Gen. {\bf 37} 9735 
(2004).

\bibitem {HM} I. Herbst, and R. Mavi, J. Phys. A: Math. Thoer. {\bf 49} 195204 
(2016). 

\bibitem {Na} K. Na, J.-W. Jung, and L.E. Reichl, J. Phys. A: Math. Theor. {\bf 43} 415304 
(2010).

\bibitem {Wi} R.G. Winter, 
Phys. Rev. {\bf 123} 1503 
(1961).

\bibitem {AgSa1} U.G. Aglietti, and P.M. Santini, 
Phys. Rev. A {\bf 89} 022111 
(2014).

\bibitem {de} R. de la Madrid, 
Nucl. Phys. A {\bf 962} 24 
(2017).

\bibitem {E} P. Exner, {\it Solvable models of resonances and decays},  in: Demuth, M., Kirsch, W. (eds) {\it Mathematical Physics, Spectral Theory and Stochastic Analysis}. Operator Theory: Advances and Applications, {\bf 232} 165-227, Birkhäuser, Basel (2013).

\bibitem {GMV} G. Garc\'\i a-Calder\'on, I. Maldonado, and J. Villavicencio, 
Phys. Rev. A {\bf 76} 012103 
(2007).

\bibitem {GL} G. Garc\'\i a-Calder\'on, and L. Chaos-Cador, 
Phys. Rev. A {\bf 90} 032109 
(2014).

\bibitem {GVHR} G. Garc\'\i a-Calder\'on, J. Villavicencio, A. Hern\'andez-Maldonado, and R. Romo, 
Phys. Rev. A {\bf 94} 022103 
(2016).

\bibitem {PVZ} M. Peshkin, A. Volya, and V. Zelevinsky, 
EPL {\bf 107} 40001 
(2014).

\bibitem {RaK} D.F. Ram{\'\i}rez Jim{\'e}nez, and N.G. Kelkar, 
J. Phys. A: Math. Theor. {\bf 52} 055201 
(2019).

\bibitem {W} S.R. Wilkinson, C.F. Bharucha, M.C. Fischer, K.W. Madison, P.R. Morrow, Q. Niu, B. Sundaram, and M.G. Raizen, 
Nature {\bf 387} 575 
(1997). 

\bibitem {Simon} B. Simon, 
Ann. Math. (NY) {\bf 97} 247 
(1973).

\bibitem {AH} J.E. Avron, and I. Herbst, Commun. Math. Phys. {\bf 52} 239 
(1977).


\end{thebibliography}
\end {document}